\documentclass[10pt,journal,twocolumns]{IEEEtran}
\usepackage{subfigure}
\usepackage{amssymb}
\usepackage{amsthm}
\usepackage{amsmath}
\usepackage{mathrsfs}
\usepackage{hyperref}
\usepackage{graphicx}
\usepackage{colortbl,dcolumn}
\usepackage{booktabs}
\usepackage{xcolor}
\usepackage{cite}
\usepackage{threeparttable}

\usepackage{algorithm}
\usepackage{algorithmic}

\usepackage{cases}

\usepackage{stmaryrd}

\newtheorem{assumption}{Assumption}
\newtheorem{definition}{Definition}
\newtheorem{lemma}{Lemma}
\newtheorem{theorem}{Theorem}

\newtheorem{corollary}{Corollary}
\newtheorem{remark}{Remark}

\newcommand{\aspref}[1]{Assumption~\ref{#1}}
\newcommand{\defref}[1]{Definition~\ref{#1}}
\newcommand{\lemref}[1]{Lemma~\ref{#1}}
\newcommand{\thmref}[1]{Theorem~\ref{#1}}

\newcommand{\corref}[1]{Corollary~\ref{#1}}
\newcommand{\figref}[1]{Fig.~\ref{#1}}

\newcommand{\secref}[1]{Section~\ref{#1}}
\newcommand{\apxref}[1]{Appendix~\ref{#1}}

\newcommand{\rekref}[1]{Remark~\ref{#1}}
\newcommand{\algref}[1]{Algorithm~\ref{#1}}

\usepackage{mathtools} 

\DeclareMathOperator*{\proj}{Proj}

\usepackage{enumerate}

\ifCLASSOPTIONcompsoc
\else
\fi
\ifCLASSINFOpdf
\else
\fi

\graphicspath{{pdfs/},{images/}}

\begin{document}
%
\title{Rethinking the Mathematical Framework and Optimality of Set-Membership Filtering}
%
%
%
%

\author{Yirui~Cong,~\IEEEmembership{Member,~IEEE,},~Xiangke~Wang,~\IEEEmembership{Senior~Member,~IEEE,}~and~Xiangyun~Zhou,~\IEEEmembership{Senior~Member,~IEEE}
\IEEEcompsocitemizethanks{\IEEEcompsocthanksitem Y.~Cong and X.~Wang are with the College of Intelligence Science and Technology, National University of Defense Technology, China.\protect\\
\indent X.~Zhou is with the Research School
of Electrical, Energy and Materials Engineering, Australian National University, Australia.\protect\\
}
}

\IEEEtitleabstractindextext{%
\begin{abstract}
Set-Membership Filter (SMF) has been extensively studied for state estimation in the presence of bounded noises with unknown statistics. Since it was first introduced in the 1960s, the studies on SMF have used the set-based description as its mathematical framework. One important issue that has been overlooked is the optimality of SMF. In this work, we put forward a new mathematical framework for SMF using concepts of uncertain variables. We first establish two basic properties of uncertain variables, namely, the law of total range (a non-stochastic version of the law of total probability) and the equivalent Bayes' rule. This enables us to put forward a general SMFing framework with established optimality. Furthermore, we obtain the optimal SMF under a non-stochastic Markov condition, which is shown to be fundamentally equivalent to the Bayes filter. Note that the classical SMF in the literature is only equivalent to the optimal SMF we obtained under the non-stochastic Markov condition. When this condition is violated, we show that the classical SMF is not optimal and it only gives an outer bound on the optimal estimation.
\end{abstract}

}

\maketitle

\IEEEdisplaynontitleabstractindextext

%
\IEEEpeerreviewmaketitle

\section{Introduction}\label{sec:Introduction}


\subsection{Motivation and Related Work}\label{sec:Motivation and Related Work}

The filtering problems in the state-space description are concerned with estimating the state information in the presence of noises, and thus are widely considered in control systems, telecommunications, navigation, and many other important fields~\cite{sarkkaS2013BOOK,ChenZ2003}.
When the statistics of the noises are known, the corresponding solution method is called stochastic filter.
A famous optimal filtering framework for Hidden Markov Models (HMMs) is the Bayes filter~\cite{sarkkaS2013BOOK,ChenZ2003,JazwinskiAH1970BOOK} which provides the complete solution to the filtering problem.
%
%
As a special case, if the noises are white Gaussian in linear systems, the corresponding Bayes filter is known as the Kalman filter~\cite{KalmanR1960}.
Note that in the Bayes filter, the white noise assumption plays an important role in supporting the optimality, since otherwise the HMM condition can hardly be guaranteed.

When the noises have unknown statistics but known ranges, the corresponding solution method is called non-stochastic filter.
In the 1960s, Witsenhausen proposed a famous filtering framework for linear systems~\cite{WitsenhausenH1966Report,WitsenhausenH1968}, which is also suitable for nonlinear systems,
%
%
known as the Set-Membership Filter (SMF).
Similarly to the Bayes filter, the SMF also has the prediction step (using the set image under system function, which becomes the Minkowski sum for linear systems) and the update step (using the set intersection).
\emph{Under this classical SMFing framework, the follow-up/existing studies focused on how to derive the exact or approximate solution for different scenarios}.
More specifically, there are mainly two types of SMFs in the literature\footnote{Interval observers~\cite{JaulinL2001BOOK,EfimovD2013,TangW2019} are not included in the SMF, since the basic idea in the update step is based on designing an observer, which is different from that of the SMFing framework discussed in this paper.}:
%
\begin{itemize}
\item   \textbf{Ellipsoidal SMF.} This type of SMFs approximates the Minkowski sum and set intersection using ellipsoidal outer bounds.
    In~\cite{SchweppeF1968}, a continuous-discrete ellipsoidal SMF was proposed to outer bound the estimate of the linear systems with two specific types of noises, which has a similar structure to the Kalman filter.
    %
    %
    With a similar system setting, both SMFing and smoothing problems were investigated in~\cite{BertsekasD1971} by solving corresponding Riccati equations.
    In~\cite{ChernouskoF1980} and~\cite{FogelE1982}, algorithms were provided for minimizing the volume of the outer bounds on the Minkowski sum and intersection of ellipsoids.
    Nevertheless, minimizing the volume can result in a very narrow ellipsoid with an unacceptably large diameter.
    Thus, the semi-axes of ellipsoids were constrained, e.g., via the trace of the matrix in the quadratic form.
    In~\cite{MaksarovD1996}, a volume-minimizing and a trace-minimizing ellipsoidal outer bounds (each described by two ellipsoids) were derived for the linear discrete-time SMF, and the description of outer bounds were generalized to multiple ellipsoids in~\cite{DurieuC2001}.
    Note that the ellipsoidal SMF is computationally cheaper but usually less accurate than the polytopic SMF discussed below.
\item   \textbf{Polytopic SMF.} This type of SMFs describes or outer bounds the prediction and the update using convex polytopes.
    Different from the ellipsoidal SMF, the polytopic SMF can derive the exact solution for linear filtering problems, because polytopes are closed under Minkowski sum and set intersection.
    Nevertheless, the complexity is unacceptable for deriving exact solutions.
    %
    %
    Noticing this fact, researchers used different subclasses of convex polytopes to give the outer bounds.
    In~\cite{ChisciL1996}, the recursive optimal bounding parallelotope algorithm was proposed to give an outer bound of the exact solution.
    %
    %
    In~\cite{CombastelC2003}, a zonotopic SMF was designed for linear discrete-time systems by using singular-value-decomposition-based approximation.
    In~\cite{AlamoT2005}, a zonotopic SMF was proposed for nonlinear discrete-time systems, where the zonotopic outer bound is derived by using convex optimization, which was improved in~\cite{AlamoT2008} by using the DC (Difference of two Convex functions) programming.
    In~\cite{LeV2013} and~\cite{CombastelC2015}, the zonotopic SMFs were given for linear systems under P-radius-based and weighted-Frobenius-norm criteria, respectively, which efficiently balanced the complexity and the accuracy of the zonotopic outer bounds.
    In~\cite{ScottJ2016}, the constrained zonotope was proposed and applied to the linear polytope-SMF, which can balance the complexity and the accuracy, and is closed under linear transformations, Minkowski sums, and set intersections.
    In~\cite{RegoB2018}, an SMF was proposed for nonlinear systems by combining the interval arithmetic and constrained zonotopes.
    In~\cite{RegoB2020}, a zonotopic SMF was studied for nonlinear systems which has advantages in handling high dimensionality.
\end{itemize}

All the above-mentioned studies on SMFs used the set-based description as its mathematical framework (see \rekref{rek:The Existing SMF Framework} for a detailed discussion).
We argue that this classical framework has the optimality issue:
for stochastic filters, we know that even with the same marginal distributions, the white noises and correlated noises in linear systems lead to different Kalman filters~\cite{JazwinskiAH1970BOOK};
for the SMFs, however, the noises with different non-stochastic correlations\footnote{Unfortunately, such non-stochastic correlations can be neither captured by the set-based description nor characterized by the statistical dependence.}
(which should result in different optimal estimations) were not distinguished;
thus, the prior studies overlooked the condition (as shown later in \aspref{asp:Unrelated Noises and Initial State}) under which the SMFs are optimal, and the optimal SMFing framework has not been rigorously established.
Departing from the classical/suboptimal set-based SMFing description, in this article, we aim to establish the optimal SMFing framework in a completely different way.



\subsection{Our Contributions}\label{sec:Our Contributions}

In this work, we put forward a new mathematical framework for SMFing, based on the concepts of uncertain variables proposed by Nair in the 2010s~\cite{NairG2013}.
Similarly to the Bayesian filtering, our filtering framework recursively derives the non-stochastic prior and posterior.
%
%
The main contributions are:

\begin{itemize}
\item   We first establish two new and fundamental properties of uncertain variables: the first one is called the \emph{law of total range}, which is a non-stochastic version of the law of total probability; the second one is the equivalent \emph{Bayes' rule for uncertain variables}.
    These properties enable us to define a new SMFing framework using the notion of uncertain variables.

\item   Most importantly, we establish \emph{an optimal SMFing framework that is more general than the well-known classical SMFing framework in the literature}.
    With this new framework, we obtain the optimal SMF under an unrelatedness assumption (to guarantee a non-stochastic Markov condition), which is shown to be fundamentally equivalent to the Bayes filter.
    We also show that \emph{the classical SMFing framework in the literature is only optimal under this Markov condition}, since it cannot capture the relatedness between the noises and the initial prior.

\item   Furthermore, we prove that when this Markoveness condition is violated, \emph{the classical SMFing gives an outer bound} on the optimal estimation.
    We also use two examples to illustrate the performance gap between the classical SMFing framework in the literature and the optimal SMFing framework proposed in this work.
\end{itemize}

\subsection{Notation}\label{sec:Paper Organization and Notation}

%

Throughout this paper, $\mathbb{R}$, $\mathbb{N}_0$, and $\mathbb{Z}_+$ denote the sets of real numbers, non-negative integers, and positive integers, respectively.
$\mathbb{R}^n$ stands for the $n$-dimensional Euclidean space.
%
%

\section{Uncertain Variables: Preliminaries and New Results}\label{sec:Uncertain Variables: Preliminaries and New Results}

In this work, the uncertainties are with known ranges but unknown probability distributions.
To model the uncertainties rigorously, we introduce the \emph{uncertain variable} proposed in~\cite{NairG2013} and derive two important properties which will constitute the foundation of the optimal SMFing framework.


\subsection{Preliminaries of Uncertain Variables}\label{sec:Preliminaries of Uncertain Variables}

Consider a sample space $\Omega$.
A measurable function $\mathbf{x}\colon \Omega \to \mathcal{X}$ from the sample space $\Omega$ to a measurable set $\mathcal{X}$ is called an uncertain variable~\cite{NairG2013}.
We define a realization of $\mathbf{x}$ as $\mathbf{x}(\omega) =: x$, and sometimes we write it as $\mathbf{x} = x$ for conciseness.

Different from random variables which can be described by probability distributions, an uncertain variable (say $\mathbf{x}$) does not have any information on the probability, but it can be described by its range $\llbracket\mathbf{x}\rrbracket$:
\begin{equation}\label{eqn:Range}
\llbracket\mathbf{x}\rrbracket := \left\{\mathbf{x}(\omega)\colon \omega \in \Omega\right\}.
\end{equation}
%

Similar to the probability distribution for multiple random variables, the range can also be defined w.r.t. multiple uncertain variables.

\begin{definition}[Joint Range, Conditional Range, Marginal Range~\cite{NairG2013}]\label{def:Joint Range, Conditional Range, Marginal Range}
Let $\mathbf{x}$ and $\mathbf{y}$ be two uncertain variables.
The joint range of $\mathbf{x}$ and $\mathbf{y}$ is
\begin{equation}\label{eqn:Joint Range}
\llbracket\mathbf{x}, \mathbf{y}\rrbracket := \left\{(\mathbf{x}(\omega), \mathbf{y}(\omega))\colon \omega \in \Omega\right\}.
\end{equation}
The conditional range of $\mathbf{x}$ given $\mathbf{y} = y$ is
\begin{equation}\label{eqn:Conditional Range}
\llbracket\mathbf{x}|y\rrbracket := \left\{\mathbf{x}(\omega)\colon \mathbf{y}(\omega) = y, \omega \in \Omega\right\} = \left\{\mathbf{x}(\omega)\colon \omega \in \Omega_{\mathbf{y}=y}\right\},
\end{equation}
where $\Omega_{\mathbf{y}=y} := \mathbf{y}^{-1}(\{y\}) = \{\omega\colon \mathbf{y}(\omega) = y, \omega \in \Omega\}$ is the preimage of $\{\mathbf{y}(\omega) = y\colon \omega \in \Omega\}$, and $\llbracket\mathbf{y}|x\rrbracket$ is defined in a similar way.
The marginal range of $\mathbf{x}$ is $\llbracket\mathbf{x}\rrbracket$ expressed by~\eqref{eqn:Range}.
\end{definition}

In analogy with the joint probability distribution, the joint range can be fully determined by the conditional and marginal ranges~\cite{NairG2013}, i.e.,
\begin{equation}\label{eqn:Joint Range Determined by Conditional and Marginal Ranges}
\llbracket\mathbf{x}, \mathbf{y}\rrbracket = \bigcup_{y \in \llbracket\mathbf{y}\rrbracket} \big(\llbracket\mathbf{x}| y\rrbracket \times \{y\}\big) = \bigcup_{x \in \llbracket\mathbf{x}\rrbracket} \big(\{x\} \times \llbracket\mathbf{y}| x\rrbracket\big),
\end{equation}
where $\times$ is the Cartesian product.

Next, we introduce the definition of unrelatedness~\cite{NairG2013}, which is a non-stochastic analogue of statistical independence.

\begin{definition}[Unrelatedness and Conditional Unrelatedness~\cite{NairG2013}]\label{def:Unrelatedness}
Uncertain variables $\mathbf{u}_1,\ldots,\mathbf{u}_r$ are unrelated if
\begin{equation}\label{eqn:Unrelatedness}
\llbracket\mathbf{u}_1,\ldots,\mathbf{u}_r\rrbracket = \llbracket\mathbf{u}_1\rrbracket \times \cdots \times \llbracket\mathbf{u}_r\rrbracket.
\end{equation}
They are conditionally unrelated given $\mathbf{v}$ if
\begin{equation}\label{eqn:Conditional Unrelatedness}
\llbracket\mathbf{u}_1,\ldots,\mathbf{u}_r|v\rrbracket = \llbracket\mathbf{u}_1|v\rrbracket \times \cdots \times \llbracket\mathbf{u}_r|v\rrbracket, \quad \forall v \in \llbracket\mathbf{v}\rrbracket.
\end{equation}
\end{definition}


If the uncertain variables are not unrelated, we say they are related.
%
%
%
Based on \defref{def:Unrelatedness}, we have the following properties for unrelatedness and conditional unrelatedness~\cite{NairG2013}:
\begin{itemize}
\item[i)]   $\mathbf{u}_1$ and $\mathbf{u}_2$ are unrelated if and only if (iff)
    \begin{equation}\label{eqn:Property of Unrelatedness}
    \llbracket\mathbf{u}_1|u_2\rrbracket = \llbracket\mathbf{u}_1\rrbracket, \quad \forall u_2 \in \llbracket\mathbf{u}_2\rrbracket.
    \end{equation}
\item[ii)]  $\mathbf{u}_1$ and $\mathbf{u}_2$ are conditionally unrelated given $v$ iff
    \begin{equation}\label{eqn:Property of Conditional Unrelatedness}
    \llbracket\mathbf{u}_1|u_2,v\rrbracket = \llbracket\mathbf{u}_1|v\rrbracket, \quad \forall (u_2,v) \in \llbracket\mathbf{u}_2,\mathbf{v}\rrbracket.
    \end{equation}
\end{itemize}

\subsection{Law of Total Range and New Bayes' Rule}\label{sec:Law of Total Range and Bayes' Rule for Uncertain Variables}

%
In this subsection, we establish two properties, namely, the law of total range and Bayes' rule for uncertain variables, as the non-stochastic counterparts of the law of total probability and Bayes' rule.
They establish a mathematical foundation of the optimal SMF which will be introduced in \secref{sec:The Optimal Filtering Framework}.

\begin{lemma}[Law of Total Range]\label{lem:Law of Total Range}
\begin{equation}\label{eqn:Law of Total Range}
\llbracket\mathbf{x}\rrbracket = \bigcup_{y \in \llbracket\mathbf{y}\rrbracket} \llbracket\mathbf{x}| y\rrbracket, \quad \llbracket\mathbf{y}\rrbracket = \bigcup_{x \in \llbracket\mathbf{x}\rrbracket} \llbracket\mathbf{y}| x\rrbracket.
\end{equation}
\end{lemma}

\begin{IEEEproof}
See \apxref{apx:Proof of lem:Law of Total Range}.
\end{IEEEproof}

\begin{figure}[ht]
\centering
\includegraphics [width=0.75\columnwidth]{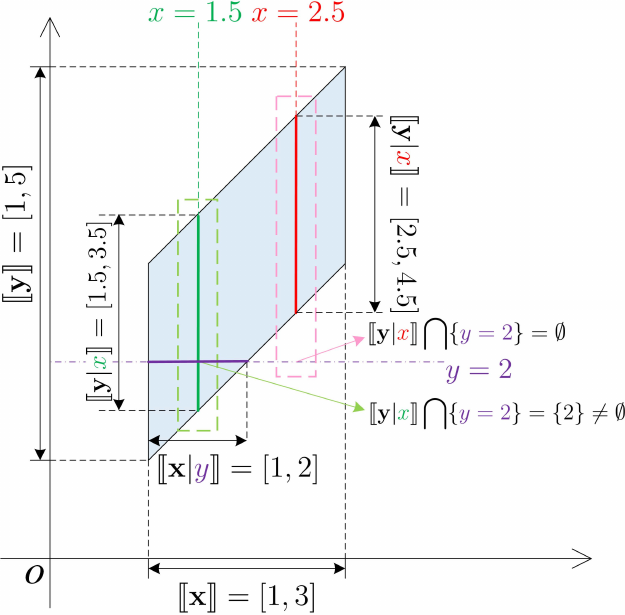}
\caption{An illustrative example of the law of total range and Bayes' rule for uncertain variables.
The prior range is $\llbracket\mathbf{x}\rrbracket = [1, 3]$, and the likelihood range is $\llbracket\mathbf{y}|x\rrbracket = \{x\} + [0, 2] = [x, x+2]$.
The law of total range~\eqref{eqn:Law of Total Range} implies $\llbracket\mathbf{y}\rrbracket = \bigcup_{x \in \llbracket\mathbf{x}\rrbracket} \llbracket\mathbf{y}|x\rrbracket = [1, 5]$ which can be verified easily in this figure.
To illustrate how Bayes' rule works, we take $\llbracket\mathbf{x}|y\rrbracket$ with $y = 2$ (the purple dash-dotted line) for an example.
When $x = 1.5$ (the green dashed line), the likelihood range marked by the green solid line segment is $[1.5,3.5]$ which has a intersection with $y = 2$, i.e., $\llbracket\mathbf{y}|x\rrbracket \bigcap \{y=2\} \neq \emptyset$.
Thus, $x = 1.5 \in \llbracket\mathbf{x}|y\rrbracket$ for $y = 2$.
When $x = 2.5$ (the red dashed line), the likelihood range marked by the red solid line segment is $[2.5,4.5]$ which has no intersections with $y = 2$, i.e., $\llbracket\mathbf{y}|x\rrbracket \bigcap \{y=2\} = \emptyset$.
Hence, $x = 2.5 \notin \llbracket\mathbf{x}|y\rrbracket$ for $y = 2$.
By applying Bayes' rule to all $x \in \llbracket\mathbf{x}\rrbracket$, the posterior range $\llbracket\mathbf{x}|y\rrbracket = [1,2]$ marked by the purple solid line segment is obtained.
}
\label{fig:Illustrative example of Bayes' rule.}
\end{figure}

The law of total range links the marginal range and the conditional range.
An illustrative example is given in \figref{fig:Illustrative example of Bayes' rule.}.
With~\eqref{eqn:Law of Total Range}, we know that $\llbracket\mathbf{x}| y\rrbracket \subseteq \llbracket\mathbf{x}\rrbracket$ which implies observations can reduce uncertainty.

\begin{lemma}[Bayes' Rule for Uncertain Variables]\label{lem:Bayes' Rule for Uncertain Variables}
\begin{equation}\label{eqn:Bayes' Rule for Uncertain Variables}
\llbracket\mathbf{x}| y\rrbracket = \left\{x\colon \llbracket\mathbf{y}| x\rrbracket \bigcap \{y\} \neq \emptyset, x \in \llbracket\mathbf{x}\rrbracket\right\}.
\end{equation}
\end{lemma}

\begin{IEEEproof}
See \apxref{apx:Proof of lem:Bayes' Rule for Uncertain Variables}.
\end{IEEEproof}

Bayes' rule for uncertain variables reflects the fundamental relationship among the prior range $\llbracket \mathbf{x} \rrbracket$, the likelihood range $\llbracket\mathbf{y}| x\rrbracket$, and the posterior range $\llbracket\mathbf{x}| y\rrbracket$.
An illustrative example is given in \figref{fig:Illustrative example of Bayes' rule.}.


\section{The Optimal Filtering Framework}\label{sec:The Optimal Filtering Framework}

Now, we model the SMFing problem in the framework of uncertain variables.
%
Consider the following nonlinear system:
\begin{align}
\mathbf{x}_{k+1} &= f_k(\mathbf{x}_k, \mathbf{w}_k),\label{eqn:State Equation}\\
\mathbf{y}_k &= g_k(\mathbf{x}_k, \mathbf{v}_k),\label{eqn:Measurement Equation}
\end{align}
for time $k \in \mathbb{N}_0$, where~\eqref{eqn:State Equation} and~\eqref{eqn:Measurement Equation} are called the state equation and the measurement equation, respectively.
The state equation describes how the system state $\mathbf{x}_k$ (with its realization $x_k \in \llbracket\mathbf{x}_k\rrbracket \subseteq \mathbb{R}^n$) changes over time, where $\mathbf{w}_k$ is the process/dynamical noise (with its realization $w_k \in \llbracket\mathbf{w}_k\rrbracket \subseteq \mathbb{R}^p$), and $f_k \colon \llbracket\mathbf{x}_k\rrbracket \times \llbracket\mathbf{w}_k\rrbracket \to \llbracket\mathbf{x}_{k+1}\rrbracket$ stands for the system transition function.
The measurement equation gives how the system state is measured, where $\mathbf{y}_k$ represents the measurement (with its realization, called observed measurement, $y_k \in \llbracket\mathbf{y}_k\rrbracket \subseteq \mathbb{R}^m$) and $\mathbf{v}_k$ (with its realization $v_k \in \llbracket\mathbf{v}_k\rrbracket \subseteq \mathbb{R}^q$) stands for the measurement noise, and $g_k \colon \llbracket\mathbf{x}_k\rrbracket \times \llbracket\mathbf{v}_k\rrbracket \to \llbracket\mathbf{y}_k\rrbracket$ is the measurement function.

Now, we define the optimality criterion for SMF and then provide the optimal SMFing framework as follows.

\begin{definition}[Optimal SMF]\label{def:Optimal SMF}
An SMF is a process that $\forall k \in \mathbb{N}_0$, it gives an estimator $X_k(y_{0:k})$ that includes all possible $x_k$ given the measurements up to $k$, i.e., $y_{0:k} := y_0,\ldots,y_k$.
An SMF is optimal if $X_k^*(y_{0:k})$ returns the smallest set such that $X_k^*(y_{0:k}) \subseteq X'_k(y_{0:k})$ holds for any $X'_k$ and $y_{0:k}$.
\end{definition}


\begin{theorem}[Optimal Set-Membership Filtering Framework]\label{thm:Optimal Set-Membership Filter}
For the system described by~\eqref{eqn:State Equation} and~\eqref{eqn:Measurement Equation}, the optimal SMF is obtained by the following steps:
\begin{itemize}
\item   \textbf{Initialization.} Set the initial prior range $\llbracket\mathbf{x}_0\rrbracket$.
\item   \textbf{Prediction.} For $k \in \mathbb{Z}_+$, given the posterior range $\llbracket\mathbf{x}_{k-1}|y_{0:k-1}\rrbracket$ in the previous time step, the prior range $\llbracket \mathbf{x}_k| y_{0:k-1}\rrbracket$ is predicted by the law of total range that
    \begin{equation}\label{eqn:Prediction - Optimal Set-Membership Filter - General Case}
    \bigcup_{x_{k-1} \in \llbracket\mathbf{x}_{k-1}|y_{0:k-1}\rrbracket}\!\!\!\!\!\!\!\!\!\!\! f_{k-1}(x_{k-1},\llbracket\mathbf{w}_{k-1}|x_{k-1}, y_{0:k-1}\rrbracket).
    \end{equation}
    %
\item   \textbf{Update.} For $k \in \mathbb{N}_0$, given the observed measurement $y_k$ and the prior range $\llbracket \mathbf{x}_k| y_{0:k-1}\rrbracket$, the posterior range $\llbracket \mathbf{x}_k| y_{0:k}\rrbracket$ is updated by Bayes' rule for uncertain variables that
    \begin{equation}\label{eqn:Update - Optimal Set-Membership Filter - General Case}
    \!\!\left\{x_k \!\in\! \llbracket\mathbf{x}_k| y_{0:k-1}\rrbracket\colon g_k(x_k, \llbracket\mathbf{v}_k|x_k, y_{0:k-1}\rrbracket) \bigcap \{y_k\} \neq \emptyset\right\}
    \end{equation}
    where we define $\llbracket\mathbf{x}_0\rrbracket := \llbracket \mathbf{x}_0| y_{0:-1}\rrbracket$ and $\llbracket\mathbf{v}_0|x_0\rrbracket = \llbracket\mathbf{v}_0|x_0, y_{0:-1}\rrbracket$ for consistency.
\end{itemize}
Note that the posterior range obtained in~\eqref{eqn:Update - Optimal Set-Membership Filter - General Case} is the optimal estimator, i.e., $X_k^*(y_{0:k}) = \llbracket \mathbf{x}_k| y_{0:k}\rrbracket$.
\end{theorem}

\begin{IEEEproof}
See~\apxref{apx:Proof of thm:Optimal Set-Membership Filter}.
\end{IEEEproof}

In general, it is not easy to obtain $\llbracket\mathbf{w}_{k-1}|x_{k-1},y_{0:k-1}\rrbracket$ and $\llbracket\mathbf{v}_k|x_k, y_{0:k-1}\rrbracket$ in \thmref{thm:Optimal Set-Membership Filter}.
They depend on how the process noises, the measurement noises, and the initial prior $\mathbf{w}_{0:k}, \mathbf{v}_{0:k},\mathbf{x}_0$ are related.
%
%
However, if the noises and the initial state are unrelated (see \aspref{asp:Unrelated Noises and Initial State}), the optimal filter is easy to derive (see \thmref{thm:Optimal Set-Membership Filter Under Assumption 1}).

\begin{assumption}[Unrelated Noises and Initial State]\label{asp:Unrelated Noises and Initial State}
$\forall k \in \mathbb{N}_0$, $\mathbf{w}_{0:k}, \mathbf{v}_{0:k},\mathbf{x}_0$ are unrelated.
\end{assumption}

\begin{theorem}[Optimal SMFing Under \aspref{asp:Unrelated Noises and Initial State}]\label{thm:Optimal Set-Membership Filter Under Assumption 1}
For the system described by~\eqref{eqn:State Equation} and~\eqref{eqn:Measurement Equation}, the optimal SMF under \aspref{asp:Unrelated Noises and Initial State}
%
is given by the following steps:
\begin{itemize}
\item   \textbf{Initialization.} Set the initial prior range $\llbracket\mathbf{x}_0\rrbracket$.
\item   \textbf{Prediction.} For $k \in \mathbb{Z}_+$, given $\llbracket\mathbf{x}_{k-1}|y_{0:k-1}\rrbracket$ derived in the previous time step $k-1$, the prior range is
    \begin{equation}\label{eqn:Prediction - Optimal Set-Membership Filter}
    \llbracket \mathbf{x}_k| y_{0:k-1}\rrbracket = f_{k-1}(\llbracket\mathbf{x}_{k-1}|y_{0:k-1}\rrbracket,\llbracket\mathbf{w}_{k-1}\rrbracket).
    \end{equation}
    %
\item   \textbf{Update.} For $k \in \mathbb{N}_0$, given the observed measurement $y_k$ and the prior range $\llbracket \mathbf{x}_k| y_{0:k-1}\rrbracket$, the posterior range is
    \begin{equation}\label{eqn:Update - Optimal Set-Membership Filter}
    \llbracket \mathbf{x}_k| y_{0:k}\rrbracket = \!\!\left[\bigcup_{v_k \in \llbracket\mathbf{v}_k\rrbracket} g_{k,v_k}^{-1} (\{y_k\})\right] \bigcap \llbracket \mathbf{x}_k| y_{0:k-1}\rrbracket,
    \end{equation}
    where $g_{k,v_k}^{-1}(\cdot)$ is the inverse map of $g_k(\cdot,v_k)$.
\end{itemize}
\end{theorem}

\begin{IEEEproof}
See \apxref{apx:Proof of thm:Optimal Set-Membership Filter Under Assumption 1}.
\end{IEEEproof}

\begin{remark}[Fundamental Equivalence Between SMF Under \aspref{asp:Unrelated Noises and Initial State} and Bayes Filter]\label{rek:Fundamental Equivalence Between SMF and Bayes Filter}
The Bayes filter~\cite{sarkkaS2013BOOK} is based on the stochastic Hidden Markov Model (HMM) with
\begin{align}
p(x_k|x_{0:k-1},y_{0:k-1}) &= p(x_k|x_{k-1}),\label{eqn:HMM - Markov Property of States}\\
p(y_k|x_{0:k},y_{0:k-1}) &= p(y_k|x_k),\label{eqn:HMM - Conditional Independence of Measuremtns}
\end{align}
where $p(a|b)$ is the conditional distribution of the random variable $\mathbf{a}$ given the realization $b = \mathbf{b}(\omega)$.
For the optimal SMF, the system described by~\eqref{eqn:State Equation} and~\eqref{eqn:Measurement Equation} under \aspref{asp:Unrelated Noises and Initial State} satisfies the following non-stochastic HMM\footnote{Equations~\eqref{eqn:Non-Stochastic HMM - Markov Property of States} and~\eqref{eqn:Non-Stochastic HMM - Conditional Independence of Measuremtns} can be proved by using \lemref{lem:Invariance of Unrelatedness under Maps} and the same technique in~\eqref{eqninpf:thm:Optimal Set-Membership Filter Under Assumption 1 - 1} of \apxref{apx:Proof of thm:Optimal Set-Membership Filter Under Assumption 1}.}:
\begin{align}
\llbracket\mathbf{x}_k|x_{0:k-1},y_{0:k-1}\rrbracket &= \llbracket\mathbf{x}_k|x_{k-1}\rrbracket,\label{eqn:Non-Stochastic HMM - Markov Property of States}\\
\llbracket\mathbf{y}_k|x_{0:k},y_{0:k-1}\rrbracket &= \llbracket\mathbf{y}_k|x_k\rrbracket.\label{eqn:Non-Stochastic HMM - Conditional Independence of Measuremtns}
\end{align}
These two HMMs are equivalent, since $p(\cdot)$ and $\llbracket\cdot\rrbracket$ describe the uncertainties for random variables and uncertain variables, respectively.
Furthermore,~\eqref{eqn:HMM - Markov Property of States} reflects the conditional independence between $\mathbf{x}_k$ and $\mathbf{x}_{0:k-2},\mathbf{y}_{0:k-1}$ given $x_{k-1}$; while~\eqref{eqn:Non-Stochastic HMM - Markov Property of States} indicates the conditional unrelatedness~\eqref{eqn:Property of Conditional Unrelatedness} between them.
Similar observation can be obtained between~\eqref{eqn:HMM - Conditional Independence of Measuremtns} and~\eqref{eqn:Non-Stochastic HMM - Conditional Independence of Measuremtns}.

In the Bayes filter, the prediction step is based on the Chapman-Kolmogorov equation, i.e., the law of total probability combined with the Markov property~\eqref{eqn:HMM - Markov Property of States} that
\begin{equation}\label{eqn:Bayes Filter - Prediction Step}
p(x_k|y_{0:k-1}) = \int p(x_k|x_{k-1}) p(x_{k-1}|y_{0:k-1}) \mathrm{d}x_{k-1}.
\end{equation}
In the optimal SMF under \aspref{asp:Unrelated Noises and Initial State}, the prediction step is given by the law of total range~\eqref{eqn:Law of Total Range} and the non-stochastic Markov property~\eqref{eqn:Non-Stochastic HMM - Markov Property of States} that\footnote{The RHS of~\eqref{eqn:SMF - Prediction Step - Fundamental Equivalence} is $f_{k-1}(\llbracket\mathbf{x}_{k-1}|y_{0:k-1}\rrbracket,\llbracket\mathbf{w}_{k-1}\rrbracket)$ as stated in \thmref{thm:Optimal Set-Membership Filter Under Assumption 1}.
But the Bayes filter does not have such an elegant expression for general nonlinear systems.}
\begin{equation}\label{eqn:SMF - Prediction Step - Fundamental Equivalence}
\llbracket \mathbf{x}_k| y_{0:k-1}\rrbracket = \bigcup_{x_{k-1} \in \llbracket\mathbf{x}_{k-1}|y_{0:k-1}\rrbracket} \llbracket\mathbf{x}_k|x_{k-1}\rrbracket.
\end{equation}

For the update steps, the Bayes filter derives the posterior distribution $p(x_k|y_{0:k})$ by Bayes' rule, while the optimal SMF gets the posterior range $\llbracket \mathbf{x}_k| y_{0:k}\rrbracket$ by Bayes' rule for uncertain variables~\eqref{eqn:Bayes' Rule for Uncertain Variables}.
\end{remark}

Further, if the system is linear, the optimal SMFing under \aspref{asp:Unrelated Noises and Initial State} is obtained in \corref{cor:Optimal Linear Set-Membership Filter}.

\begin{corollary}\label{cor:Optimal Linear Set-Membership Filter}
For the linear system described by
\begin{align}
\mathbf{x}_{k+1} &= A \mathbf{x}_k + B \mathbf{w}_k,\label{eqn:State Equation - Linear System}\\
\mathbf{y}_k &= C \mathbf{x}_k + D \mathbf{v}_k,\label{eqn:Measurement Equation - Linear System}
\end{align}
where $A \in \mathbb{R}^{n\times n}$, $B \in \mathbb{R}^{n\times p}$, $C \in \mathbb{R}^{m\times n}$, and $D \in \mathbb{R}^{m\times q}$,
the optimal SMF under \aspref{asp:Unrelated Noises and Initial State} has the following steps:
\begin{itemize}
\item   \textbf{Initialization.} Set the initial prior range $\llbracket\mathbf{x}_0\rrbracket$.
\item   \textbf{Prediction.} For $k \in \mathbb{Z}_+$, the prior range is
    \begin{equation}\label{eqn:Prediction - Optimal Linear Set-Membership Filter}
    \llbracket \mathbf{x}_k| y_{0:k-1}\rrbracket = A \llbracket\mathbf{x}_{k-1}|y_{0:k-1}\rrbracket \oplus B \llbracket\mathbf{w}_{k-1}\rrbracket,
    \end{equation}
    where $\oplus$ stands for the Minkowski sum\footnote{Given two sets $\mathcal{S}_1$ and $\mathcal{S}_2$ in Euclidean space, the Minkowski sum of $\mathcal{S}_1$ and $\mathcal{S}_2$ is $\mathcal{S}_1 \oplus \mathcal{S}_2 = \{s_1 + s_2\colon s_1 \in \mathcal{S}_1, s_2 \in \mathcal{S}_2\}$.}.
\item   \textbf{Update.} For $k \in \mathbb{N}_0$, given $y_k$, the posterior range is
    \begin{equation}\label{eqn:Update - Optimal Linear Set-Membership Filter}
    \llbracket \mathbf{x}_k| y_{0:k}\rrbracket = \mathcal{X}_k(C, y_k, D\llbracket\mathbf{v}_k\rrbracket) \bigcap \llbracket \mathbf{x}_k| y_{0:k-1}\rrbracket,
    \end{equation}
    where we define $\llbracket\mathbf{x}_0\rrbracket := \llbracket \mathbf{x}_0| y_{0:-1}\rrbracket$ for consistency, and $\mathcal{X}_k(C, y_k, D\llbracket\mathbf{v}_k\rrbracket) = \{x_k\colon y_k = C x_k + D v_k, v_k \in \llbracket\mathbf{v}_k\rrbracket\}$.
    %
    %
\end{itemize}
\end{corollary}

\begin{remark}[The Existing SMFing Framework]\label{rek:The Existing SMF Framework}
The classical SMFing framework in the literature is under the set-based description:
e.g.,~\cite{SchweppeF1968,ChisciL1996,ScottJ2016} for linear filters and~\cite{LeongP2016,RegoB2020} for nonlinear filters.
Specifically:
\begin{itemize}
\item   In~\cite{SchweppeF1968}, the classical SMFing framework was applied [see equations~(9) and~(11) therein] to linear systems, and an ellipsoidal outer bound was proposed.
\item   In~\cite{ChisciL1996}, the classical SMFing framework was also considered [see equations~(5) and~(6) therein] for linear systems, and a paralleltopic outer bound was given.
\item   In~\cite{ScottJ2016}, the classical SMFing framework was also employed [see equation~(32) therein] for linear systems, and the exact solution or outer bounds can be given by the proposed constrained zonotopes.
\item   In~\cite{RegoB2020}, the classical SMFing framework was also used [see equations~(2) and~(3) therein] for nonlinear systems, and an efficient constrained zonotopic SMF was designed based on two new methods (i.e., the mean value and first-order Taylor extensions).
\item   In~\cite{LeongP2016}, the classical SMFing framework was also taken into account [see Lemma~1 therein] for nonlinear systems, and an approximate solution was derived by proposing a novel particle filter.
\end{itemize}
Note that all these prior works require \aspref{asp:Unrelated Noises and Initial State} to hold.
However, when \aspref{asp:Unrelated Noises and Initial State} is violated, the property of non-stochastic HMM described by~\eqref{eqn:Non-Stochastic HMM - Markov Property of States} and~\eqref{eqn:Non-Stochastic HMM - Conditional Independence of Measuremtns} can hardly be guaranteed.
Without this property, the classical SMFing framework is not optimal any more, i.e., it cannot give the exact set of all possible states determined by the optimal SMFing framework in \thmref{thm:Optimal Set-Membership Filter}.
\end{remark}




Although the classical SMFing framework does not give the optimal solution for state estimation when \aspref{asp:Unrelated Noises and Initial State} is violated, the following theorem tells that it is still useful in giving a more conservative estimate.

\begin{theorem}[Outer Bound]\label{thm:Outer Boundedness}
Let $\llbracket\mathbf{x}_k^*|y_{0:k}\rrbracket$ and $\llbracket\mathbf{x}_k|y_{0:k}\rrbracket$ be the posterior ranges derived by \thmref{thm:Optimal Set-Membership Filter} and \thmref{thm:Optimal Set-Membership Filter Under Assumption 1}, respectively.
Then, $\llbracket\mathbf{x}_k^*|y_{0:k}\rrbracket \subseteq \llbracket\mathbf{x}_k|y_{0:k}\rrbracket$ holds.
\end{theorem}

\begin{IEEEproof}
See \apxref{apx:Proof of thm:Outer Boundedness}.
\end{IEEEproof}

Furthermore, a class of systems with state-and-process-noise relatedness can be converted to non-stochastic HMMs with the following model-modification method.

\begin{remark}[Relatedness Cancellation]\label{rek:Relatedness Cancellation}
For related states $\mathbf{x}_{0:k}$ and process noises $\mathbf{w}_{0:k}$, let $\mathbf{z}_k = [\mathbf{x}_k^{\mathrm{T}}, \mathbf{w}_k^{\mathrm{T}}]^{\mathrm{T}}$ ($k \in \mathbb{N}_0$) be the new state, if the system described by~\eqref{eqn:State Equation} and~\eqref{eqn:Measurement Equation} can be rewritten as
\begin{align}
\begin{bmatrix}
\mathbf{x}_{k+1} \\
\mathbf{w}_{k+1}
\end{bmatrix}=
\mathbf{z}_{k+1} &= \bar{f}_k (\mathbf{z}_k) =
\begin{bmatrix}
\bar{f}_k^{(x)} (\mathbf{z}_k)\\
\bar{f}_k^{(w)} (\mathbf{z}_k)
\end{bmatrix},\label{eqn:Modified State Equation}\\
\mathbf{y}_k &= \bar{g}_k(\mathbf{z}_k, \mathbf{v}_k) = g_k(\mathbf{x}_k, \mathbf{v}_k),\label{eqn:Modified Measurement Equation}
\end{align}
for $\forall k \in \mathbb{N}_0$, where $\bar{\mathbf{v}}_{0:k}$ and $\mathbf{z}_0$ are unrelated (i.e., \aspref{asp:Unrelated Noises and Initial State} holds).
Then, the optimal SMF can be obtained by directly applying \thmref{thm:Optimal Set-Membership Filter Under Assumption 1} to the modified system described by~\eqref{eqn:Modified State Equation} and~\eqref{eqn:Modified Measurement Equation}.
\end{remark}

Nevertheless, the relatedness cancellation method in \rekref{rek:Relatedness Cancellation} cannot deal with all kinds of relatedness, such as inequality-type relatedness (e.g., $\mathbf{w}_k + \mathbf{w}_{k-1} \leq 1$) and the related noises in \secref{sec:Nonlinear System with Related Process and Measurement Noises}.

\section{Numerical Examples}\label{sec:Numerical Examples}

In this section, we illustrate the performance gap between the optimal SMFing framwork (in \thmref{thm:Optimal Set-Membership Filter}) and the classical framework (equivalent to \thmref{thm:Optimal Set-Membership Filter Under Assumption 1}) through two examples.


\subsection{System with Related Process and Measurement Noises}\label{sec:Nonlinear System with Related Process and Measurement Noises}

Consider the nonlinear system described by
\begin{align}
\mathbf{x}_{k+1} &= \sin(\mathbf{x}_k) + \mathbf{x}_k + \mathbf{w}_k,\label{eqn:Simulation - Nonlinear System - State Equation}\\
\mathbf{y}_k &= \mathbf{v}_k \mathbf{x}_k,\label{eqn:Simulation - Nonlinear System - Measurement Equation}
\end{align}
where $\llbracket\mathbf{x}_0\rrbracket = [0, 1]$, $\llbracket\mathbf{w}_k\rrbracket = [0, 1]$, and $\llbracket\mathbf{v}_k\rrbracket = [1, 2]$.
$\forall k \in \mathbb{N}_0$, $\mathbf{w}_{0:k},\mathbf{v}_{0:k},\mathbf{x}_0$ are unrelated, except that the process noise $\mathbf{w}_{k-1}$ and the multiplicative measurement noise $\mathbf{v}_k$ satisfy $\llbracket\mathbf{v}_k|w_{k-1}\rrbracket = \left[\max\{1, 1.8-w_{k-1}\},~2 - w_{k-1}\right]$ ($k \in \mathbb{Z}_+$).

If we ignore this relatedness, \algref{alg:Classical SMF} will give the exact solution for classical SMFing (in \thmref{thm:Optimal Set-Membership Filter Under Assumption 1}), where Line~\ref{line:Prediction Step} gives the prediction step and Line~\ref{line:Update Step} provides the update step.

\begin{algorithm}
\begin{footnotesize}
\caption{Optimal Algorithm Under Classical Framework}\label{alg:Classical SMF}
\begin{algorithmic}[1]
\STATE  \textbf{Initialization:} $\llbracket\mathbf{x}_0\rrbracket = [a_0^-, b_0^-] = [0, 1]$; \COMMENT{\textbf{Comments:} $\llbracket\mathbf{x}_k|y_{0:k}\rrbracket = [a_k, b_k]$ ($k \in \mathbb{Z}_+$), $\llbracket\mathbf{x}_k|y_{0:k-1}\rrbracket = [a_k^-, b_k^-]$ ($k \in \mathbb{N}_0$).}
\LOOP
    \IF {$k > 0$}
        \STATE  $a_k^- = \sin(a_{k-1}) + a_{k-1}$, $b_k^- = \sin(b_{k-1}) + b_{k-1} + 1$;\label{line:Prediction Step}
    \ENDIF
    \STATE  $[a_k, b_k] = [\min\{0.5 y_k,y_k\},\max\{0.5 y_k,y_k\}] \bigcap [a_k^-, b_k^-]$;\label{line:Update Step}
    \STATE  $k = k + 1$;
\ENDLOOP
\end{algorithmic}
\end{footnotesize}
\end{algorithm}

\begin{algorithm}
\begin{footnotesize}
\caption{Approximation of the Optimal SMF}\label{alg:Random-Sample-Based Approximation of the Optimal Filter}
\begin{algorithmic}[1]
\STATE  \textbf{Initialization:} $\llbracket\mathbf{x}_0\rrbracket = [0, 1]$, $N = 10000$;\label{line:Initialization} \COMMENT{\textbf{Comments:} $N$ is the number of random samples for $\llbracket\mathbf{x}_k^*|y_{0:k}\rrbracket = [a_k^*, b_k^*]$ ($k \in \mathbb{Z}_+$).}
\LOOP\label{line:Loop - Start}
    \IF {$k = 0$}\label{line:k = 0}
        \STATE  $[a_0^*,b_0^*] = \llbracket\mathbf{x}_0^*|y_0\rrbracket \leftarrow \llbracket\mathbf{x}_0|y_0\rrbracket$ in \algref{alg:Classical SMF};
    \ELSIF  {$k > 0$}\label{line:k > 0}
        \STATE  $i = 1$;
        \WHILE  {$i \leq N$}
        \STATE  $x_{k-1} \leftarrow U(a_{k-1}^*,b_{k-1}^*)$ and $w_{k-1} \leftarrow U(0, 1)$; \COMMENT{\textbf{Comments:} $x \leftarrow U(a,b)$ means $x$ is a realization of a random variable uniformly distributed in $[a, b]$.}
        \STATE  $x_k = \sin(x_{k-1}) + x_{k-1} + w_{k-1}$;
        \IF {($x_k \neq 0~\&\&~\frac{y_k}{x_k} \in \llbracket\mathbf{v}_k|w_{k-1}\rrbracket$)~$||$~($x_k = y_k = 0$)}
            \STATE  $x_k^{(i)} \leftarrow x_k$, $i = i + 1$;
        \ENDIF
        \ENDWHILE
        \STATE  $a_k^* = \min_i \{x_k^{(i)}\}$, $b_k^* = \max_i \{x_k^{(i)}\}$;
    \ENDIF
    \STATE  $k = k + 1$;
\ENDLOOP\label{line:Loop - End}
\end{algorithmic}
\end{footnotesize}
\end{algorithm}


Now, we design the optimal SMF from \thmref{thm:Optimal Set-Membership Filter}, and its state range is denoted by $\llbracket\mathbf{x}^*|\cdot\rrbracket$ to be distinguished from the ranges in \algref{alg:Classical SMF}.
For $k = 0$, the posterior range $\llbracket\mathbf{x}_0^*|y_0\rrbracket$ is identical to that derived in \algref{alg:Classical SMF}, as $\llbracket\mathbf{x}_0^*\rrbracket = \llbracket\mathbf{x}_0\rrbracket$.
For $k > 1$, assume $\llbracket\mathbf{x}_{k-1}^*|y_{0:k-1}\rrbracket := [a_{k-1}^*,b_{k-1}^*]$ has already been derived at $k - 1$.
Since $\mathbf{w}_{k-1}$ is only related to $\mathbf{v}_k$, we have $\llbracket\mathbf{w}_{k-1}|x_{k-1},y_{0:k}\rrbracket = \llbracket\mathbf{w}_{k-1}\rrbracket$ in~\eqref{eqn:Prediction - Optimal Set-Membership Filter - General Case} of the prediction step.
Similarly, we have $\llbracket\mathbf{v}_k|x_k, y_{0:k-1}\rrbracket = \llbracket\mathbf{v}_k|x_k\rrbracket$ in~\eqref{eqn:Update - Optimal Set-Membership Filter - General Case} of the update step.
However, we would not use~\eqref{eqn:Prediction - Optimal Set-Membership Filter - General Case} to obtain $\llbracket\mathbf{x}_k^*|y_{0:k-1}\rrbracket$ directly, because in the update step, $\llbracket\mathbf{v}_k|x_k\rrbracket$ is not explicit which cannot help to derive $\llbracket\mathbf{x}_k^*|y_{0:k}\rrbracket$.
Instead, we can rewrite~\eqref{eqn:Prediction - Optimal Set-Membership Filter - General Case} and~\eqref{eqn:Update - Optimal Set-Membership Filter - General Case} as
%
\begin{multline}\label{eqn:Update - Optimal Set-Membership Filter - General Case - Rewritten}
\Big\{x_k = f_{k-1}(x_{k-1},w_{k-1})\colon g_{k,x_k}^{-1}(\{y_k\}) \in \llbracket\mathbf{v}_k|w_{k-1}\rrbracket,\\ x_{k-1} \in \llbracket\mathbf{x}_{k-1}^*|y_{0:k-1}\rrbracket, w_{k-1} \in \llbracket\mathbf{w}_{k-1}\rrbracket\Big\},
\end{multline}
where $g_{k,x_k}^{-1}(\cdot)$ is the inverse map of $g_k(x_k,\cdot)$.
From~\eqref{eqn:Update - Optimal Set-Membership Filter - General Case - Rewritten}, we can derive the posterior range by the following steps:
for each $x_{k-1} \in \llbracket\mathbf{x}_{k-1}^*|y_{0:k-1}\rrbracket$ and $w_{k-1} \in \llbracket\mathbf{w}_{k-1}\rrbracket$, we calculate $x_k = \sin(x_{k-1}) + x_{k-1} + w_{k-1}$ via~\eqref{eqn:Simulation - Nonlinear System - State Equation};
if $g_{k,x_k}^{-1}(\{y_k\}) \in \llbracket\mathbf{v}_k|w_{k-1}\rrbracket$, then $x_k = f_{k-1}(x_{k-1},w_{k-1})$ is a possible state in $\llbracket\mathbf{x}_k^*|y_{0:k}\rrbracket$.
With all such possible $x_k$, we get the posterior range $\llbracket\mathbf{x}_k^*|y_{0:k}\rrbracket$.
The Monte Carlo technique can be employed to approximate the posterior range (see \algref{alg:Random-Sample-Based Approximation of the Optimal Filter}).

\begin{figure}[ht]
\centering
\includegraphics [width=0.83\columnwidth]{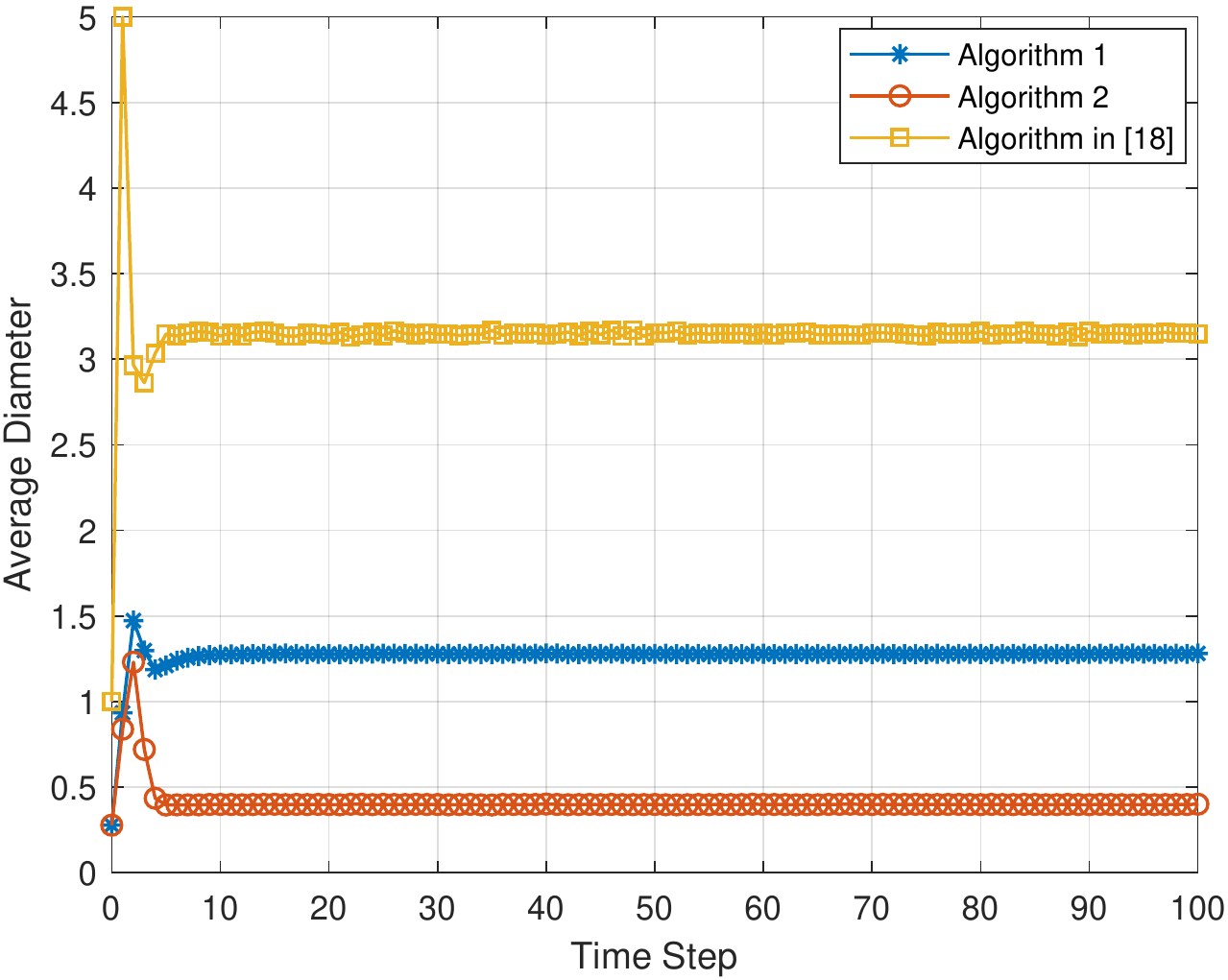}
\caption{Comparison of \algref{alg:Classical SMF}, \algref{alg:Random-Sample-Based Approximation of the Optimal Filter}, and the algorithm in~\cite{AlamoT2005}.
The diameter of each algorithm is averaged over $10000$ simulation runs.
The averaged execution times in each time step for \algref{alg:Classical SMF}, \algref{alg:Random-Sample-Based Approximation of the Optimal Filter}, and the algorithm in~\cite{AlamoT2005} are $1.144 \times 10^{-5}$s, $3.181 \times 10^{-3}$s and $1.194 \times 10^{-5}$s, respectively, where the simulation is conducted by using Matlab 2019b on a laptop with Intel Core i7-7700HQ@2.80GHz CPU.
}
\label{fig:Interval-length ratio.}
\end{figure}

\figref{fig:Interval-length ratio.} shows the average diameters of the estimates in \algref{alg:Classical SMF}, \algref{alg:Random-Sample-Based Approximation of the Optimal Filter}, and the algorithm in~\cite{AlamoT2005}, respectively.\footnote{The probability distributions of uncertain variables $\mathbf{x}_0,\mathbf{w}_{0:k},\mathbf{v}_{0:k}$ can be arbitrary for simulations.
In \secref{sec:Numerical Examples}, these uncertain variables are set to be uniformly distributed in their ranges/conditional ranges.}
We can see that the optimal SMF in \algref{alg:Random-Sample-Based Approximation of the Optimal Filter} performs the best, which corroborates our theoretical results.

\subsection{Linear System with Identical Process Noise}\label{sec:Linear System with Identical Process Noise}

Consider the linear system described by
\begin{align}
\mathbf{x}_{k+1} &=
\begin{bmatrix}
1 & 1 \\
0 & 1
\end{bmatrix}
\mathbf{x}_k +
\begin{bmatrix}
0.5 \\
1
\end{bmatrix}
\mathbf{w},\label{eqn:Simulation - Linear System - State Equation}\\
\mathbf{y}_k &=
\begin{bmatrix}
1 & 0
\end{bmatrix}\mathbf{x}_k + \mathbf{v}_k,\label{eqn:Simulation - Linear System - Measurement Equation}
\end{align}
where $\llbracket\mathbf{x}_0\rrbracket = [-10, 10]\times[-10, 10]$, $\llbracket\mathbf{w}\rrbracket = [-1, 1]$, and $\llbracket\mathbf{v}_k\rrbracket = [-1, 1]$.
$\forall k \in \mathbb{N}_0$, $\mathbf{x}_0,\mathbf{w},\mathbf{v}_{0:k}$ are unrelated.

If we replace $\mathbf{w}$ with $\mathbf{w}_k$ and assume \aspref{asp:Unrelated Noises and Initial State} holds, the classical SMFing is \corref{cor:Optimal Linear Set-Membership Filter} with
\begin{equation}\label{eqn:System Paramters - Linear System}
A =
\begin{bmatrix}
1 & 1 \\
0 & 1
\end{bmatrix},\quad
B =
\begin{bmatrix}
0.5\\
1
\end{bmatrix},\quad
C =
\begin{bmatrix}
1 & 0
\end{bmatrix},\quad
D = 1.
\end{equation}
We employ the Projection-Based (PB) method in~\cite{ShammaJ1999} to give the estimate $\llbracket\mathbf{x}_k|y_{0:k}\rrbracket$ exactly, labeled as PB-SMF 1.

Now we design the optimal SMF using \rekref{rek:Relatedness Cancellation}, and the modified system with $\mathbf{z}_k = [\mathbf{x}_k^{\mathrm{T}}, \mathbf{w}_k]^{\mathrm{T}} = [\mathbf{x}_k^{(1)}, \mathbf{x}_k^{(2)}, \mathbf{w}_k]^{\mathrm{T}}$ is
\begin{align}
\mathbf{z}_{k+1} &=
\begin{bmatrix}
1 & 1 & 0.5\\
0 & 1 & 1\\
0 & 0 & 1
\end{bmatrix}
\mathbf{z}_k,\label{eqn:Simulation - Linear System - State Equation - Rewritten}\\
\mathbf{y}_k &=
\begin{bmatrix}
1 & 0 & 0
\end{bmatrix}\mathbf{z}_k + \mathbf{v}_k,\label{eqn:Simulation - Linear System - Measurement Equation - Rewritten}
\end{align}
which gives the optimal filter, labeled as PB-SMF 2, when \corref{cor:Optimal Linear Set-Membership Filter} is applied.
Similarly to \secref{sec:Nonlinear System with Related Process and Measurement Noises}, we denote $\llbracket\mathbf{x}_k^*|y_{0:k}\rrbracket$ as the optimal posterior range, which can be derived by the projection of $\llbracket\mathbf{z}_k|y_{0:k}\rrbracket$ to the $x^{(1)}x^{(2)}$-plane.

\begin{figure}
\centering

\subfigure[]{\includegraphics [width=0.64\columnwidth]{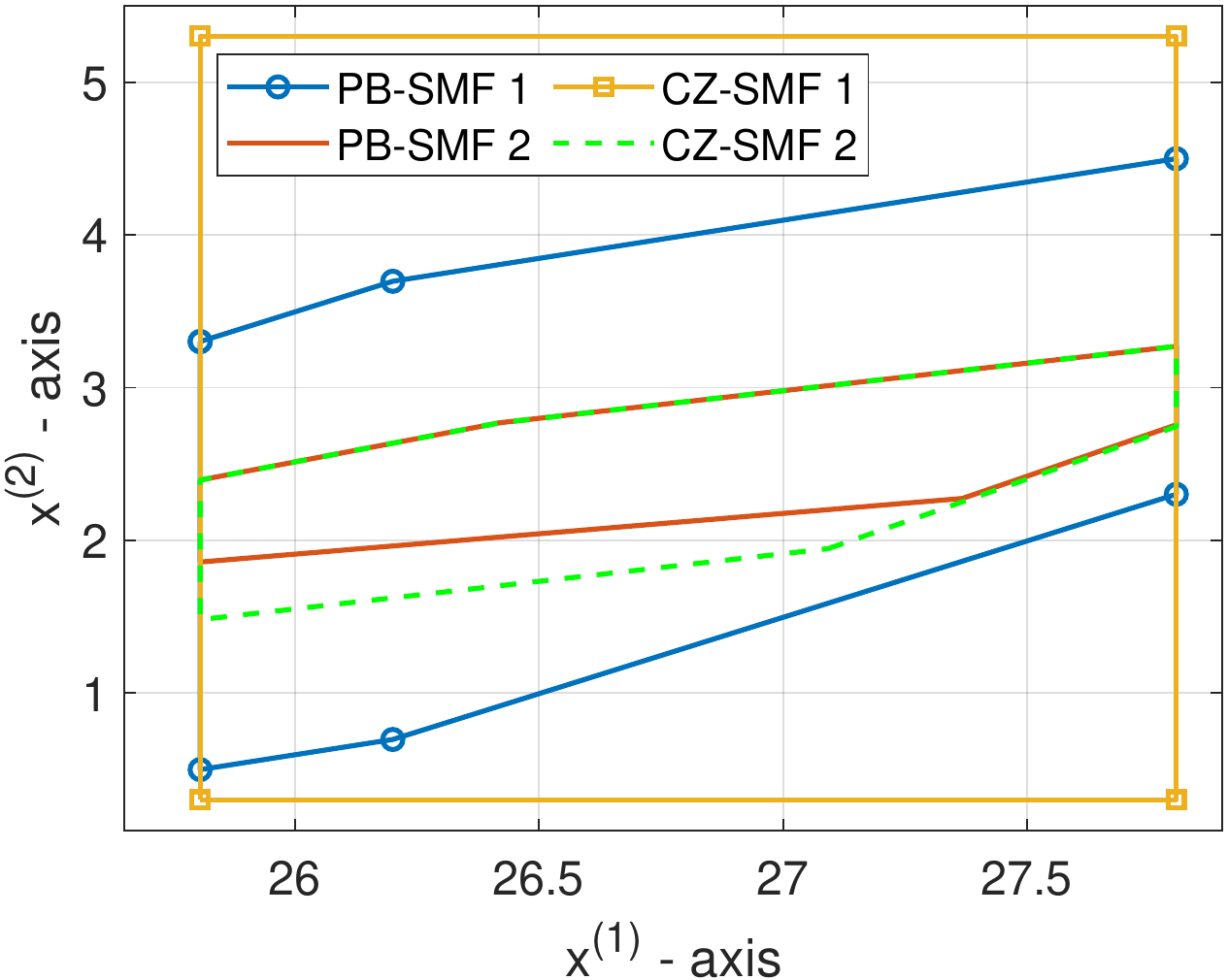}\label{fig:Posterior range comparison - B}}\\
\subfigure[]{\includegraphics [width=0.64\columnwidth]{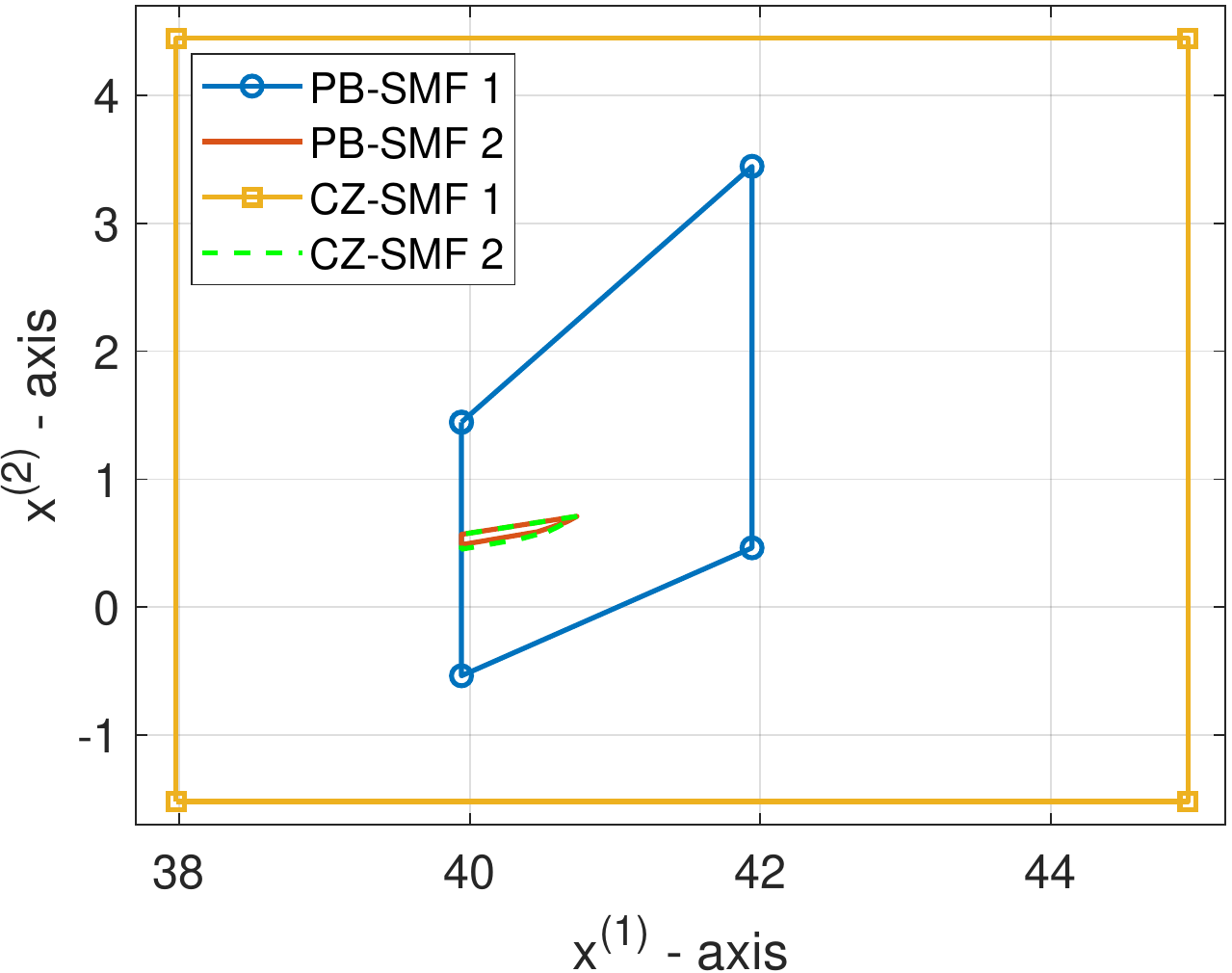}\label{fig:Posterior range comparison - C}}

\caption{Comparisons of the PB-SMF and the CZ-SMF under the classical and the optimal SMFing frameworks at:
(a) $k = 10$,
(b) $k = 20$.
CZ-SMF 1 (under the classical framework) and CZ-SMF 2 (under the optimal framework) are simulated with the help of CORA 2020 toolbox~\cite{AlthoffM2020Manual}, where the degree-of-freedom order and the number of constraints are $0$ and $5$, respectively.
}\label{fig:Posterior range comparison.}
\end{figure}

\figref{fig:Posterior range comparison.} shows the performance gap between the (optimal) PB-SMFs under the optimal and classical frameworks.
We can see that $\llbracket\mathbf{x}_{10}^*|y_{0:10}\rrbracket$ (based on PB-SMF2) is smaller than $\llbracket\mathbf{x}_{10}|y_{0:10}\rrbracket$ (based on PB-SMF1) at $k = 10$ in \figref{fig:Posterior range comparison - B}, and the area ratio is $27.1\%$;
finally, $\llbracket\mathbf{x}_{10}^*|y_{0:20}\rrbracket$ becomes much smaller than $\llbracket\mathbf{x}_{20}|y_{0:20}\rrbracket$ at $k = 20$ in \figref{fig:Posterior range comparison - C}, and the area ratio is $1.04\%$, which means approximately $99\%$ of the estimated range by the classical SMFing is excluded by the optimal SMFing.
Besides, \figref{fig:Posterior range comparison.} also presents the gaps between the Constrained Zonotopic SMF (CZ-SMF)~\cite{ScottJ2016} under the optimal and classical frameworks.
The area ratios are $19.0\%$ and $0.174\%$ for $k = 10$ and $k = 20$, respectively.

\section{Conclusion}\label{sec:Conclusion}

In this work, we have studied the optimal SMFing problem for discrete-time systems.
Based on the uncertain variables, we have put forward an optimal SMFing framework.
Then, we have obtained the optimal SMF under non-stochastic Markov condition, and revealed the fundamental equivalence between the SMF and the Bayes filter.
We have also shown that the classical SMF in the literature must rely on the non-stochastic Markov condition to guarantee optimality.
When the Markovness is violated, the classical SMF is not optimal and can only provide an outer bound on the optimal estimation.
%

\appendices

\section{Proof of \lemref{lem:Law of Total Range}}\label{apx:Proof of lem:Law of Total Range}

We only prove $\llbracket\mathbf{x}\rrbracket = \bigcup_{y \in \llbracket\mathbf{y}\rrbracket} \llbracket\mathbf{x}| y\rrbracket$, and the proof for $\llbracket\mathbf{y}\rrbracket = \bigcup_{x \in \llbracket\mathbf{x}\rrbracket} \llbracket\mathbf{y}| x\rrbracket$ is similar.
From~\eqref{eqn:Conditional Range}, we have $\bigcup_{y \in \llbracket\mathbf{y}\rrbracket} \llbracket\mathbf{x}| y\rrbracket = \bigcup_{y \in \llbracket\mathbf{y}\rrbracket} \left\{\mathbf{x}(\omega)\colon \omega \in \Omega_{\mathbf{y}=y}\right\}
\stackrel{(a)}{=} \left\{\mathbf{x}(\omega)\colon \omega \in  \Omega\right\} = \llbracket\mathbf{x}\rrbracket$,
where $(a)$ is from $\bigcup_{y \in \llbracket\mathbf{y}\rrbracket} \mathbf{x}(\Omega_{\mathbf{y}=y}) = \mathbf{x}(\bigcup_{y \in \llbracket\mathbf{y}\rrbracket} \Omega_{\mathbf{y}=y}) = \mathbf{x}(\Omega)$.
Thus,~\eqref{eqn:Law of Total Range} holds.\hfill$\blacksquare$

\section{Proof of \lemref{lem:Bayes' Rule for Uncertain Variables}}\label{apx:Proof of lem:Bayes' Rule for Uncertain Variables}

Firstly, we define $\llbracket\mathbf{x}, y\rrbracket$ as $\{(\mathbf{x}(\omega), \mathbf{y}(\omega))\colon \mathbf{y}(\omega)
= y, \omega \in \Omega\} = \{(\mathbf{x}(\omega), \mathbf{y}(\omega))\colon \omega \in \Omega_{\mathbf{y}=y}\}$.
With~\eqref{eqn:Conditional Range}, we have
\begin{equation}\label{eqn:Total Range with One Variable Fixed - Expressed by Conditional Range}
\llbracket\mathbf{x}, y\rrbracket = \llbracket\mathbf{x}| y\rrbracket \times \{y\},
\end{equation}
and conversely we have
\begin{equation}\label{eqn:Conditional Range - Expressed by Total Range with One Variable Fixed}
\llbracket\mathbf{x}| y\rrbracket = \proj_{(x,y)\mapsto x} \llbracket\mathbf{x}, y\rrbracket,
\end{equation}
where $\proj_{(x,y)\mapsto x}(\cdot)$ is a projection from the space w.r.t. $(x, y)$ to the subspace w.r.t. $x$ that $\proj_{(x,y)\mapsto x}(\mathcal{S}_x, \mathcal{S}_y) = \mathcal{S}_x$ for sets $\mathcal{S}_x$ and $\mathcal{S}_y$.

Secondly, we prove the following equation holds
\begin{equation}\label{eqninpf:Bayes' Rule for Uncertain Variables - 1}
\llbracket\mathbf{x}, y\rrbracket = \llbracket\mathbf{x}, \mathbf{y}\rrbracket \bigcap \left(\llbracket\mathbf{x}\rrbracket \times \{y\}\right).
\end{equation}
With the RHS of the first equality in~\eqref{eqn:Joint Range Determined by Conditional and Marginal Ranges}, the RHS of~\eqref{eqninpf:Bayes' Rule for Uncertain Variables - 1} can be rewritten as
\begin{equation}\label{eqninpf:Bayes' Rule for Uncertain Variables - 2}
\begin{split}
&\Big[\bigcup_{y' \in \llbracket\mathbf{y}\rrbracket} \big(\llbracket\mathbf{x}| y\rrbracket \times \{y'\}\big)\Big] \bigcap \left(\llbracket\mathbf{x}\rrbracket \times \{y\}\right)\\
=&\bigcup_{y' \in \llbracket\mathbf{y}\rrbracket} \left[\big(\llbracket\mathbf{x}| y'\rrbracket \times \{y'\}\big) \bigcap \left(\llbracket\mathbf{x}\rrbracket \times \{y\}\right)\right]\\
\stackrel{(a)}{=}&\bigcup_{y \in \llbracket\mathbf{y}\rrbracket} \left[\big(\llbracket\mathbf{x}| y'\rrbracket \bigcap \llbracket\mathbf{x}\rrbracket\big) \times \big(\{y'\} \bigcap \{y\}\big)\right]\\
\stackrel{(b)}{=}&\llbracket\mathbf{x}| y\rrbracket \times \{y\} \stackrel{(c)}{=} \llbracket\mathbf{x}, y\rrbracket,
\end{split}
\end{equation}
where $(a)$ follows $(\mathcal{S}_1 \times \mathcal{S}_2) \bigcap (\mathcal{S}_3 \times \mathcal{S}_4) = (\mathcal{S}_1 \bigcap \mathcal{S}_3) \times (\mathcal{S}_2 \bigcap \mathcal{S}_4)$ for sets $\mathcal{S}_1,\ldots,\mathcal{S}_4$.
Equality $(b)$ is established by~\eqref{eqn:Law of Total Range} (which implies $\llbracket\mathbf{x}| y'\rrbracket \subseteq \llbracket\mathbf{x}\rrbracket$) and $\mathcal{S} \times \emptyset = \emptyset$ for set $\mathcal{S}$.
Then, $(c)$ follows from~\eqref{eqn:Total Range with One Variable Fixed - Expressed by Conditional Range}.

Thirdly, we prove a projection-based version of Bayes' rule
\begin{equation}\label{eqn:Projection-Based Version of Bayes' rule}
\llbracket\mathbf{x}| y\rrbracket = \proj_{(x,y)\mapsto x}\Bigg(\bigcup_{x \in \llbracket\mathbf{x}\rrbracket} \bigg[\{x\} \times \left(\llbracket\mathbf{y}| x\rrbracket \bigcap \{y\}\right)\bigg]\Bigg).
\end{equation}
With~\eqref{eqninpf:Bayes' Rule for Uncertain Variables - 1} and the RHS of the second equality in~\eqref{eqn:Joint Range Determined by Conditional and Marginal Ranges}, we get
\begin{equation}\label{eqninpf:Bayes' Rule for Uncertain Variables - 3}
\begin{split}
\llbracket\mathbf{x}, y\rrbracket &= \bigg[\bigcup_{x \in \llbracket\mathbf{x}\rrbracket} \big(\{x\} \times \llbracket\mathbf{y}| x\rrbracket\big)\bigg] \bigcap \left(\llbracket\mathbf{x}\rrbracket \times \{y\}\right)\\
&= \bigcup_{x \in \llbracket\mathbf{x}\rrbracket} \left[\big(\{x\} \times \llbracket\mathbf{y}| x\rrbracket\big) \bigcap \left(\llbracket\mathbf{x}\rrbracket \times \{y\}\right)\right]\\
&= \bigcup_{x \in \llbracket\mathbf{x}\rrbracket} \left[\big(\{x\} \bigcap \llbracket\mathbf{x}\rrbracket\big) \times \big( \llbracket\mathbf{y}| x\rrbracket \bigcap \{y\} \big)\right]\\
&= \bigcup_{x \in \llbracket\mathbf{x}\rrbracket} \left[\{x\} \times \big(\llbracket\mathbf{y}| x\rrbracket \bigcap \{y\}\big)\right].
\end{split}
\end{equation}
By~\eqref{eqn:Conditional Range - Expressed by Total Range with One Variable Fixed} and~\eqref{eqninpf:Bayes' Rule for Uncertain Variables - 3},~\eqref{eqn:Projection-Based Version of Bayes' rule} is obtained.

Finally, we prove that~\eqref{eqn:Bayes' Rule for Uncertain Variables} and~\eqref{eqn:Projection-Based Version of Bayes' rule} are equivalent.
Let $\mathcal{T}_1$ and $\mathcal{T}_2$ denote the RHS of~\eqref{eqn:Projection-Based Version of Bayes' rule} and the RHS of~\eqref{eqn:Bayes' Rule for Uncertain Variables}, respectively.
$\forall x' \in \mathcal{T}_1$, we have $\llbracket\mathbf{y}| x'\rrbracket \bigcap \{y\} \neq \emptyset$, since otherwise $\{x'\} \times \big(\llbracket\mathbf{y}| x'\rrbracket \bigcap \{y\}\big) = \emptyset$ which means $x' \notin \mathcal{T}_1$.
Observing that $x' \in \llbracket\mathbf{x}\rrbracket$, we get $x' \in \mathcal{T}_2$, and thus $\mathcal{T}_1 \subseteq \mathcal{T}_2$.
Conversely, $\forall x'' \in \mathcal{T}_2$, we have $x'' \in \llbracket\mathbf{x}\rrbracket$ and $\llbracket\mathbf{y}| x''\rrbracket \bigcap \{y\} \neq \emptyset$.
Hence, $\{x''\} \times \big(\llbracket\mathbf{y}| x''\rrbracket \bigcap \{y\}\big) \neq \emptyset$ holds with $x'' \in \llbracket\mathbf{x}\rrbracket$ which implies $x'' \in \mathcal{T}_1$ and therefore $\mathcal{T}_2 \subseteq \mathcal{T}_1$.
Combining it with $\mathcal{T}_1 \subseteq \mathcal{T}_2$, we get $\mathcal{T}_1 = \mathcal{T}_2$.\hfill$\blacksquare$

\section{Proof of \thmref{thm:Optimal Set-Membership Filter}}\label{apx:Proof of thm:Optimal Set-Membership Filter}

We divide the proof of \thmref{thm:Optimal Set-Membership Filter} into two parts, the prediction step and the update step.

For the prediction step, the law of total range in~\eqref{eqn:Law of Total Range} gives
\begin{equation}\label{eqninpf:thm - Optimal Set-Membership Filter - Prediction - 1}
\llbracket \mathbf{x}_k| y_{0:k-1}\rrbracket = \bigcup_{x_{k-1} \in \llbracket\mathbf{x}_{k-1}|y_{0:k-1}\rrbracket} \llbracket\mathbf{x}_k|x_{k-1}, y_{0:k-1}\rrbracket.
\end{equation}
From~\eqref{eqn:State Equation}, the following holds
\begin{equation}\label{eqninpf:thm - Optimal Set-Membership Filter - Prediction - 2}
\begin{split}
\llbracket\mathbf{x}_k|x_{k-1}, y_{0:k-1}\rrbracket &= \llbracket f(\mathbf{x}_{k-1}, \mathbf{w}_{k-1})|x_{k-1}, y_{0:k-1}\rrbracket\\
&= \llbracket f(x_{k-1}, \mathbf{w}_{k-1})|x_{k-1}, y_{0:k-1}\rrbracket\\
&\stackrel{(a)}{=} f(x_{k-1}, \llbracket\mathbf{w}_{k-1}|x_{k-1}, y_{0:k-1}\rrbracket),
\end{split}
\end{equation}
where $(a)$ follows from~\eqref{eqn:Conditional Range} that
\begin{equation}\label{eqninpf:thm - Optimal Set-Membership Filter - Prediction - 3}
\begin{split}
&\llbracket f(x_{k-1}, \mathbf{w}_{k-1})|x_{k-1}, y_{0:k-1}\rrbracket\\
=& \{f(x_{k-1}, \mathbf{w}_{k-1}(\omega))\colon \omega \in \Omega_{\mathbf{x}_{k-1},\mathbf{y}_{0:k-1}=x_{k-1},y_{0:k-1}}\}\\
=& \{f(x_{k-1}, w_{k-1})\colon w_{k-1} \in \llbracket\mathbf{w}_{k-1}|x_{k-1}, y_{0:k-1}\rrbracket\}\\
=& f(x_{k-1}, \llbracket\mathbf{w}_{k-1}|x_{k-1}, y_{0:k-1}\rrbracket).
\end{split}
\end{equation}
Combining~\eqref{eqninpf:thm - Optimal Set-Membership Filter - Prediction - 1} with~\eqref{eqninpf:thm - Optimal Set-Membership Filter - Prediction - 2}, we get~\eqref{eqn:Prediction - Optimal Set-Membership Filter - General Case}.

For the update step, we prove it with Bayes' rule for uncertain variables.
From~\eqref{eqn:Bayes' Rule for Uncertain Variables}, we have
\begin{equation}\label{eqninpf:thm - Optimal Set-Membership Filter - Update - 1}
\llbracket \mathbf{x}_k| y_{0:k}\rrbracket \!=\! \Big\{x_k \!\in \llbracket\mathbf{x}_k| y_{0:k-1}\rrbracket\colon \llbracket\mathbf{y}_k| x_k, y_{0:k-1}\rrbracket \bigcap \{y_k\} \neq \emptyset\Big\}.
\end{equation}
Similarly to dealing with $\llbracket\mathbf{x}_k|x_{k-1}, y_{0:k-1}\rrbracket$ in~\eqref{eqninpf:thm - Optimal Set-Membership Filter - Prediction - 2}, we have $\llbracket\mathbf{y}_k| x_k, y_{0:k-1}\rrbracket = g_k(x_k, \llbracket\mathbf{v}_k|x_k, y_{0:k-1}\rrbracket)$.
Thus, the RHS of~\eqref{eqninpf:thm - Optimal Set-Membership Filter - Update - 1} can be rewritten as~\eqref{eqn:Update - Optimal Set-Membership Filter - General Case}.

By \defref{def:Joint Range, Conditional Range, Marginal Range}, the set of all possible $x_k$ given $y_{0:k}$ is exact the posterior range $\llbracket\mathbf{x}_k|y_{0:k}\rrbracket$.
Therefore, $X_k^*(y_{0:k}) = \llbracket\mathbf{x}_k|y_{0:k}\rrbracket$ which satisfies the condition $X_k^*(y_{0:k}) \subseteq X'_k(y_{0:k})$ holds for any $X'_k$ and $y_{0:k}$ in \defref{def:Optimal SMF}.\hfill$\blacksquare$

\section{Proof of \thmref{thm:Optimal Set-Membership Filter Under Assumption 1}}\label{apx:Proof of thm:Optimal Set-Membership Filter Under Assumption 1}

Before start, we need the following two lemmas.

\begin{lemma}[Function of Conditional Range]\label{lem:Function of Conditional Range}
Given uncertain variables $\mathbf{u}_1, \mathbf{u}_2$ and map $h$, $\llbracket h(\mathbf{u}_1)|u_2\rrbracket = h(\llbracket \mathbf{u}_1|u_2\rrbracket)$ holds.
%
\end{lemma}

\begin{IEEEproof}
$\llbracket h(\mathbf{u}_1)|u_2\rrbracket = \{h(\mathbf{u}_1(\omega))\colon \omega \in \Omega_{\mathbf{u}_2 = u_2}\} = h(\{\mathbf{u}_1(\omega)\colon \omega \in \Omega_{\mathbf{u}_2 = u_2}\}) = h(\llbracket \mathbf{u}_1|u_2\rrbracket)$.
\end{IEEEproof}

\begin{lemma}[Invariance of Unrelatedness under Maps]\label{lem:Invariance of Unrelatedness under Maps}
If $\mathbf{u}_1$ and $\mathbf{u}_2$ are unrelated, $h_1(\mathbf{u}_1)$ and $h_2(\mathbf{u}_2)$ are unrelated, i.e.,
\begin{equation}\label{eqn:Invariance of Unrelatedness under Maps}
\llbracket h_1(\mathbf{u}_1)|h_2(u_2)\rrbracket = \llbracket h_1(\mathbf{u}_1)\rrbracket, \quad \forall u_2 \in \llbracket\mathbf{u}_2\rrbracket.
\end{equation}
\end{lemma}

\begin{IEEEproof}
By \lemref{lem:Function of Conditional Range}, the LHS and RHS of equation~\eqref{eqn:Invariance of Unrelatedness under Maps} can be written as $h_1(\llbracket \mathbf{u}_1|h_2(u_2)\rrbracket)$ and $h_1(\llbracket \mathbf{u}_1\rrbracket)$, respectively.
Since a sufficient condition to $h_1(\llbracket \mathbf{u}_1|h_2(u_2)\rrbracket) = h_1(\llbracket \mathbf{u}_1\rrbracket)$ is
\begin{equation}\label{eqninpf:Invariance of Unrelatedness under Maps - A Sufficient Condition}
\llbracket \mathbf{u}_1|h_2(u_2)\rrbracket = \llbracket \mathbf{u}_1\rrbracket,
\end{equation}
we need to prove that~\eqref{eqninpf:Invariance of Unrelatedness under Maps - A Sufficient Condition} holds for $u_2 \in \llbracket\mathbf{u}_2\rrbracket$.

$\forall u_2 \in \llbracket\mathbf{u}_2\rrbracket$, we have $\llbracket \mathbf{u}_1|u_2\rrbracket = \{\mathbf{u}_1(\omega)\colon \omega \in \Omega_{\mathbf{u}_2 = u_2}\}$ and $\llbracket \mathbf{u}_1|h_2(u_2)\rrbracket = \{\mathbf{u}_1(\omega)\colon \omega \in \Omega_{h_2(\mathbf{u}_2) = h_2(u_2)}\}$.
As $\Omega_{h_2(\mathbf{u}_2) = h_2(u_2)} = \mathbf{u}_2^{-1} \circ h_2^{-1}(\{h_2(u_2)\})$ and $h_2^{-1} (\{h_2(u_2)\}) \supseteq \{u_2\}$, we get $\Omega_{h_2(\mathbf{u}_2) = h_2(u_2)} \supseteq \Omega_{\mathbf{u}_2 = u_2}$ which implies $\llbracket \mathbf{u}_1|h_2(u_2)\rrbracket \supseteq \llbracket \mathbf{u}_1|u_2\rrbracket$.
Thus~\eqref{eqninpf:Invariance of Unrelatedness under Maps - A Sufficient Condition} is established by
\begin{equation}\label{eqninpf:Invariance of Unrelatedness under Maps - 3}
\llbracket \mathbf{u}_1\rrbracket \supseteq \llbracket \mathbf{u}_1|h_2(u_2)\rrbracket \supseteq \llbracket \mathbf{u}_1|u_2\rrbracket \stackrel{(a)}{=} \llbracket \mathbf{u}_1\rrbracket,
\end{equation}
where $(a)$ follows from the fact that $\forall u_2 \in \llbracket\mathbf{u}_2\rrbracket$, $\llbracket \mathbf{u}_1|u_2\rrbracket = \llbracket \mathbf{u}_1\rrbracket$ for unrelated $\mathbf{u}_1$ and $\mathbf{u}_2$ [see~\eqref{eqn:Property of Unrelatedness}].
Therefore,~\eqref{eqn:Invariance of Unrelatedness under Maps} holds, and combining it with~\eqref{eqn:Property of Unrelatedness}, we know that $h_1(\mathbf{u}_1)$ and $h_2(\mathbf{u}_2)$ are also unrelated.
\end{IEEEproof}

Now we prove the prediction and update steps in \thmref{thm:Optimal Set-Membership Filter Under Assumption 1}, respectively.
In the prediction step, for $\llbracket\mathbf{w}_{k-1}|x_{k-1},y_{0:k-1}\rrbracket$ in~\eqref{eqn:Prediction - Optimal Set-Membership Filter - General Case}, we know that the collection of $\mathbf{x}_{k-1},\mathbf{y}_{0:k-1}$ is a function of
$\mathbf{w}_{0:k-2}, \mathbf{v}_{0:k-1}, \mathbf{x}_0 =: {\boldsymbol \varpi}_{k-1}$,
i.e., $(\mathbf{x}_{k-1},\mathbf{y}_{0:k-1}) =: \xi({\boldsymbol \varpi}_{k-1})$.
By \aspref{asp:Unrelated Noises and Initial State}, $\mathbf{w}_{k-1}$ and ${\boldsymbol \varpi}_{k-1}$ are unrelated.
Thus, applying \lemref{lem:Invariance of Unrelatedness under Maps}, we get
\begin{equation}\label{eqninpf:thm:Optimal Set-Membership Filter Under Assumption 1 - 1}
\llbracket\mathbf{w}_{k-1}|x_{k-1},y_{0:k-1}\rrbracket = \llbracket\mathbf{w}_{k-1}|\xi(\varpi_{k-1})\rrbracket = \llbracket\mathbf{w}_{k-1}\rrbracket,
\end{equation}
where $\varpi_{k-1}$ is the realization of ${\boldsymbol \varpi}_{k-1}$.
With~\eqref{eqninpf:thm:Optimal Set-Membership Filter Under Assumption 1 - 1},~\eqref{eqn:Prediction - Optimal Set-Membership Filter - General Case} becomes~\eqref{eqn:Prediction - Optimal Set-Membership Filter}.

In the update step, we can use a similar technique in~\eqref{eqninpf:thm:Optimal Set-Membership Filter Under Assumption 1 - 1} to obtain $\llbracket\mathbf{v}_k|x_k, y_{0:k-1}\rrbracket = \llbracket\mathbf{v}_k\rrbracket$.
Then,~\eqref{eqn:Update - Optimal Set-Membership Filter - General Case} becomes
\begin{equation}\label{eqninpf:thm:Optimal Set-Membership Filter Under Assumption 1 - 2}
\begin{split}
&\!\!\!\!\left\{x_k \in \llbracket\mathbf{x}_k| y_{0:k-1}\rrbracket\colon g_k(x_k, \llbracket\mathbf{v}_k\rrbracket) \bigcap \{y_k\} \neq \emptyset\right\}\\
&\!\!\!\!\stackrel{(a)}{=}\!\!\bigcup_{v_k \in \llbracket\mathbf{v}_k\rrbracket} \left\{x_k \in \llbracket\mathbf{x}_k| y_{0:k-1}\rrbracket\colon  \{x_k\} = g_{k,v_k}^{-1} (\{y_k\})\right\}\\
&\!\!\!\!=\!\!\bigcup_{v_k \in \llbracket\mathbf{v}_k\rrbracket} \left[g_{k,v_k}^{-1} (\{y_k\}) \bigcap \llbracket \mathbf{x}_k|y_{0:k-1}\rrbracket\right]
= \mathrm{RHS~of~\eqref{eqn:Update - Optimal Set-Membership Filter}}.
\end{split}
\end{equation}
where $(a)$ is from $g_k(x_k, \llbracket\mathbf{v}_k\rrbracket) = \bigcup_{v_k \in \llbracket\mathbf{v}_k\rrbracket} \{g_k(x_k, v_k)\}$
and the fact that $\{g_k(x_k, v_k)\} \bigcap \{y_k\} \neq \emptyset$ iff $\{x_k\} = g_{k,v_k}^{-1} (\{y_k\})$, in which $g_{k,v_k}^{-1} (\{y_k\}) = \{x_k\colon g_k(x_k, v_k) = y_k\}$.
%
%
\hfill$\blacksquare$

\section{Proof of \thmref{thm:Outer Boundedness}}\label{apx:Proof of thm:Outer Boundedness}

In the initialization step, $\llbracket\mathbf{x}_0^*\rrbracket = \llbracket\mathbf{x}_0\rrbracket$ holds.
In the update step at $k = 0$, since~\eqref{eqn:Law of Total Range} implies $\llbracket\mathbf{x}|y\rrbracket \subseteq \llbracket\mathbf{x}\rrbracket$, we have $\llbracket\mathbf{v}_0|x_0\rrbracket \subseteq \llbracket\mathbf{v}_0\rrbracket$ in~\eqref{eqn:Update - Optimal Set-Membership Filter - General Case}.
Thus,
\begin{multline}\label{eqninpf:thm:Outer Boundedness - 1}
\llbracket\mathbf{x}_0^*|y_0\rrbracket = \left\{x_0 \!\in\! \llbracket\mathbf{x}_0^*\rrbracket\colon g_0(x_0, \llbracket\mathbf{v}_0|x_0\rrbracket) \bigcap \{y_0\} \neq \emptyset\right\}\\
\subseteq \left\{x_0 \!\in\! \llbracket\mathbf{x}_0^*\rrbracket\colon g_0(x_0, \llbracket\mathbf{v}_0\rrbracket) \bigcap \{y_0\} \neq \emptyset\right\} \stackrel{(a)}{=} \llbracket\mathbf{x}_0|y_0\rrbracket,
\end{multline}
where $(a)$ follows from~\eqref{eqn:Update - Optimal Set-Membership Filter} and~\eqref{eqninpf:thm:Optimal Set-Membership Filter Under Assumption 1 - 2}.
%
%
Similarly, in the prediction step at $k = 1$, we have $\llbracket\mathbf{w}_1|x_0,y_0\rrbracket \subseteq \llbracket\mathbf{w}_1\rrbracket$, which implies $\llbracket\mathbf{x}_1^*|y_0\rrbracket \subseteq \llbracket\mathbf{x}_1|y_0\rrbracket$.
Proceeding forward, we get $\llbracket\mathbf{x}_k^*|y_{0:k}\rrbracket \subseteq \llbracket\mathbf{x}_k|y_{0:k}\rrbracket$ for $k \in \mathbb{N}_0$.\hfill$\blacksquare$

\bibliographystyle{IEEEtran}

\bibliography{ROOSMF}
\end{document}